\documentclass[conference]{IEEEtran}
\usepackage{graphicx}
\usepackage{amssymb}
\usepackage{amsmath}
\usepackage{mathrsfs}
\usepackage{float}
\usepackage{epstopdf}
\usepackage{mathrsfs}

\renewcommand{\baselinestretch}{0.91}

\newcommand{\ignore}[1]{}

\begin{document}
\renewcommand\thepage{}
\title{\LARGE \bf Application of Data Collected by Endpoint Detection and Response Systems for Implementation of a Network Security System based on Zero Trust Principles and the EigenTrust Algorithm}
\author{Nitesh Kumar, Gaurav S. Kasbekar and D. Manjunath}

\maketitle

{\renewcommand{\thefootnote}{} \footnotetext{N. Kumar, G. S. Kasbekar and D. Manjunath are with the Department of Electrical Engineering, Indian Institute of Technology Bombay, Mumbai, India. Their email addresses are niteshkumarx@iitb.ac.in, gskasbekar@ee.iitb.ac.in  and dmanju@ee.iitb.ac.in, respectively.}}

\begin{abstract}
Traditionally, security systems for enterprises have implicit access based on strong cryptography, authentication and key sharing, wherein access control is based on Role Based Access Control (RBAC), in which roles such as manager, accountant and so on provide a way of deciding a subject's authority. However, years of post-attack analysis on enterprise networks has shown that a majority of times, security breaches occur intentionally or accidently due to implicitly trusted people of an enterprise itself. Zero Trust Architecture works on the principle of never granting trust implicitly, but rather continuously evaluating the trust parameters for each resource  access request and has a strict, but not rigid, set of protocols for access control of a subject to resources. Endpoint Detection and Response (EDR) systems are tools that collect a large number of attributes in and around  machines within an enterprise network to have close visibility into sophisticated intrusion. In our work, we seek to deploy EDR systems and build trust algorithms using tactical provenance analysis, threshold cryptography and reputation management to continuously record data, evaluate trust of a subject, and simultaneously analyze them against a database of known threat vectors to provide conditional access control. However, EDR tools generate a high volume of data that leads to false alarms, misdetections and correspondingly a high backlog of tasks that makes it infeasible, which is addressed using tactical provenance analysis and information theory.
\end{abstract}

\section{Introduction}
\label{sec:introduction}
Years of post-attack analysis on enterprise networks has shown that a majority of times, security breaches occur intentionally or accidently due to implicitly trusted people of an enterprise itself~\cite{RF:forrester_research}. The reason for this is that most enterprises have traditionally been following Role Based Access Control (RBAC)~\cite{RF:zta_attributes}, wherein roles such as manager, accountant and so on provide a way of deciding a subject's authority. Due to this, a person with a high designation and authority is, by default, allowed access to sensitive data. Zero Trust Architecture~\cite{RF:zta_nist} works on the principle of never granting trust implicitly, but rather continuously evaluating the trust parameters for each resource  access request and has a strict, but not rigid, set of protocols for access control of a subject to resources (e.g., software applications, network access, storage, processor time)~\cite{RF:zta_nist}.

Also, Advanced Persistent Threats (APTs)~\cite{RF:tactical_provenance},~\cite{RF:apt} are a type of adversaries that lurk in an enterprise network for a long time to extend their reach  before initiating devastating attacks~\cite{RF:tactical_provenance}. Endpoint Detection and Response (EDR)~\cite{RF:tactical_provenance} systems are tools that collect a large number of attributes in and around  machines within an enterprise network to have close visibility into sophisticated intrusion~\cite{RF:tactical_provenance}. 

Access in Zero Trust Architecture is based on Attribute Based Access Control (ABAC)~\cite{RF:zta_attributes}, wherein a decision to grant or deny a subject's request is strictly based on enterprise specific attributes (e.g., user-id, user-authentication-pass, time-stamp-of-access, external-network-id, function-calls, system-calls) of subjects. Operationally, this translates into significantly shrinking implicit trust zones~\cite{RF:zta_nist} and increasing granularity~\cite{RF:zta_attributes} of access control and access to resources. An implicit trust zone represents an area where all the entities are trusted to at least the level of the last policy enforcement point gateway~\cite{RF:zta_nist}.  Attributes periodically collected using EDR tools may vary as per their weight in the function of a trust score and its trustworthiness~\cite{RF:zta_nist}, based on which access to a resource is granted. A trust score function can also vary from network to network and can be altered as per requirements as some weak attributes for an enterprise can be strong attributes for another~\cite{RF:access_control_nist}.

Two key issues need to be addressed while implementing EDR. The first is the huge volume of data generated by EDR systems, which becomes difficult to process in real-time~\cite{RF:tactical_provenance}. The effect is even more when granularity of control is increased while EDR systems are deployed at all subjects~\cite{RF:tactical_provenance}. The second  is a high rate of false alarms~\cite{RF:tactical_provenance}. Despite this, we cannot avoid continuous recording of attributes as threats such as APT require a continous evaluation to track down their increasing reach in the network~\cite{RF:tactical_provenance}. Now, since trust scores are to be continually evaluated, long term behavioral data from EDR systems needs to be persistently maintained; thus, we need a method for significantly reducing the size of the log records for long term evaluation and improving the trust assignment algorithm~\cite{RF:zta_nist},~\cite{RF:zta_attributes}. 
Furthermore, the threat perception evaluated from the activity needs to have high accuracy. Later, we can build a threat score assignment algorithm. 

Tactical Provenance Analysis (TPA)~\cite{RF:tactical_provenance} inspects the temporal and causal ordering of threat alerts within the Tactical Provenance Graph (TPG)~\cite{RF:tactical_provenance} to identify sequences of APT attack actions and later does graph reduction~\cite{RF:tactical_provenance} to shrink the path of the graph that is not usable in threat detection. We seek to build  a trust algorithm by applying Tactical Provenance Analysis~\cite{RF:tactical_provenance},  Threshold Cryptography~\cite{RF:threshold_crypto_1999},~\cite{RF:threshold_crypto_1994},~\cite{RF:shamir_adi_threshold},~\cite{RF:shared_auth_sig},~\cite{RF:rsa_threshold} and the EigenTrust Algorithm~\cite{RF:eigen_trust_reputation_management} to build a robust Zero Trust Architecture for enterprise networks. 

Next, recall that under threshold cryptography, a public key is used to encrypt the message, and the associated private key is shared among the participants~\cite{RF:threshold_crypto_1999}. E.g., when we are following an $(n, t+1)$ threshold cryptography scheme, then it allows $n$ computers to share the ability to perform a cryptographic operation by any $t +1$ parties jointly, but not by $t$ parties~\cite{RF:threshold_crypto_1999}. We also seek to use the EigenTrust Algorithm, which is based on the notion of transitive trust~\cite{RF:eigen_trust_reputation_management}, which means that if a peer $p$ trusts any peer $q$, then it would also trust the peers trusted by peer $q$~\cite{RF:eigen_trust_reputation_management}. This also takes our work towards the possibility of decentralization of the Zero Trust Architecture. 

The rest of the paper is organized as follows. In Section~\ref{sec:related:work}, we provide more details about the previous works~\cite{RF:tactical_provenance},~\cite{RF:zta_nist},~\cite{RF:eigen_trust_reputation_management},~\cite{RF:shamir_adi_threshold}, which are most relevant to our work. We provide our system design in Section~\ref{sec:system:design} and a discussion, conclusions and directions for future research in Section~\ref{sec:conclusion}. 

\section{Related Work}
\label{sec:related:work}

In~\cite{RF:tactical_provenance}, it has been shown that tactical provenance analysis can be applied to system logs to parse host events into provenance graphs~\cite{RF:tactical_provenance} that describe the totality of system execution and facilitate causal analysis of system activities. Those causal dependencies are then encoded into a TPG. The work states that attacks like APTs can be prevented if we know the tactics, techniques and procedures~\cite{RF:tactical_provenance} in which a user is executing events in an order, and later we can match them against a knowledge base like MITRE ATT\&CK~\cite{RF:tactical_provenance}. A remarkable reduction of false alarms is claimed in this work with a reduced size of log records that also helps in context generation~\cite{RF:tactical_provenance} from logs, which is otherwise quite a laborious process and requires a lot of manual effort and time.

The work in~\cite{RF:zta_nist} explains policy making using ABAC and its importance in the implementation of ZTA with a clear and  unambiguous attribute based approach~\cite{RF:zta_nist}. It divides ZTA into two planes-- control plane and data plane; all the information needed for access control of the system is transferred through the control plane and the resource data is transferred through the data plane. The Policy Decision Point (PDP)~\cite{RF:zta_nist} is the system within the control plane where all the policies are built, deployed and updated (an attempt is made to build algorithms adaptively as per future requirements). The inputs of PDP are through various trust vectors and external resources~\cite{RF:zta_nist}. Policy Enforcement Point (PEP)~\cite{RF:zta_nist} is the system within the control plane which actually passes grant access signals after the access request has been validated by the PDP. 

In the seminal work~\cite{RF:shamir_adi_threshold}, threshold cryptography was  introduced by designing a way to divide data $D$ into $n$ pieces in such a way that $D$ is easily reconstructable from any $z$ pieces, where $2 \leq z < n$, but even complete knowledge of $z-1$ pieces reveals absolutely no information about $D$~\cite{RF:shamir_adi_threshold}. This technique enables the construction of robust key management schemes for cryptographic systems.

In~\cite{RF:eigen_trust_reputation_management},  the idea of reputation management in a decentralized system by means of transitive trust of peers was introduced. The basic idea is that for  a peer $p$ to have trust in a third party peer $r$, we can check if  peer $q$ is trusted by peer $p$ and peer $q$ trusts the third party $r$, and if yes, then $p$ can consider peer $r$ as trustworthy~\cite{RF:eigen_trust_reputation_management}.

\section{System Design for a Zero Trust Architecture Based Enterprise}
\label{sec:system:design}

\subsection{Defining a Family of Attributes to Form Trust Vectors}
\label{sec:attributes:define}
We define a family of running trust scores for each triplet (user $i$, device $j$, resource $k$) that can be used to control access to resources in a ZTA based system. This triplet gives a unique identity to the access request. Trust scores~\cite{RF:zta_nist} in addition can be a function of various attributes~\cite{RF:zta_attributes} and their corresponding weights. Some examples are as follows: average amount of time for external network access, duration of flash drive usage, entry timestamp, number of input/ output operations, number of privilege escalation attempts, number of malicious file accesses, most frequent external network access, number of function calls, number of system calls, exit timestamp and so on. On EDR systems, these can be stored as database records after applying TPG and graph reduction techniques such as Minimally Sufficient Skeleton~\cite{RF:tactical_provenance} to reduce the overall number of records, say $N_R$, and cardinality, say $N_C$, of records. Here, by cardinality, we mean the number of attributes in the record.

\subsection{Log Compression using Information Theory}
\label{sec:log:compression}
The size of log records can be drastically reduced by using some standard coding schemes in information theory such as Huffman coding~\cite{RF:info_theory}. We can choose to uniquely code repetitive complete set or subset of attributes as given in Section~\ref{sec:attributes:define}, as per its probability, say $p_i$, of occurrence and cardinality, say $N_i$. Our challenge is to give higher probability records with a lower size of label that can be done using Huffman coding tree~\cite{RF:info_theory}. Overall, we need to reduce the average length, say $A_L$, of the records for better storage and reducing the size of logs; note that:
\[
\mathbf{A_L}=\min \sum_i{ \mathbf{p}_{i} \times N_i}.
\]

\subsection{Database Architecture for EDR Summary}
\label{sec:database:architecture}
In our framework, the tables required for database storage should correspond to a finite number of attributes. Also, enough tables should be used to declare a system activity, system state and trust level for each triplet (user $i$, device $j$, resource $k$). We consider the number of attributes to be finite for the complexity of the algorithm to be bounded. In addition, for different purposes, we may need multiple tables in the database such as one for  EDR recording keeping with a limited pool of a week of log record, another for compressed and reduced data set for prolonged storage and future analysis, and one additional table for access control~\cite{RF:access_control_nist},~\cite{RF:atribute_based_access_control_nist},~\cite{RF:access_control_ztn} and credentials. Our database system would be having the flexibility to add or remove  tables, attributes and perform database normalization~\cite{RF:database} as per requirements. 

\subsection{Distributed Algorithm Design for the Computation of Trust Scores and Control Access}
\label{sec:algorithm:trustscore}
In order to discard the overhead and single point of computation and bottleneck with dependency, we want to keep our algorithm distributed so that even if a few nodes fail, then the other systems keep running. In~\cite{RF:eigen_trust_reputation_management}, a peer to peer transitive trust mechanism called Eigentrust that is not centralized was defined. We are studying various low complexity algorithm design techniques reliability to provide improved security features by explicitly assuming that at a given time instant, some of the nodes in an enterprise network in which EDR tools are installed may be compromised and/ or failed. We seek to mitigate the effect of delays from the time instant of infection of nodes until they are detected by security analysts using EDR tools, by designing and implementing a secure and distributed method to limit the damage to the enterprise network caused by compromised nodes until their infection is detected and patched.

\subsection{Caching and Updating of Trust Scores}
\label{sec:caching}
In order to maintain a running trust score for each request triplet (user $i$, device $j$, resource $k$), we use an EDR system to keep updating the trust scores in real-time or keep a refreshing time of trust scores as small as possible, say five minutes. There will be a dedicated centralized server in the network that will be responsible for keeping variables like $\mbox{Last-Known-Recent-Trust-Score}$ and $\mbox{Cached-Score}$ for each request triplet. Cached scores will be available only for requests identified as frequent and will be stored on another faster storage device like a solid state drive with a superior clock speed. We can adopt any optimal cache memory replacement schemes like Least Recently Used (LRU)~\cite{RF:cache_memory_hamacher}, which replaces cache blocks that are not frequently requested with one having a higher frequency of request. To minimize the real time computation delay, we do not encourage computing the trust score at the time of access requirements, but instead, the $\mbox{Last-Known-Recent-Trust-Score}$ should always be pre-computed and available with a $\mbox{MAX-Refresh-Time}$ of five minutes. At the time of access control demand, we can get the trust score with a single query operation rather than computing it which may first be a hit on $\mbox{Cached-Score}$ on miss the score will be looked into dedicated centralized server for the $\mbox{Last-Known-Recent-Trust-Score}$.

\subsection{Combining Reputation Management and Zero Trust Principles to build Trust Algorithms}
\label{sec:combining:reputation:tacticalprovenance:threshold_cryptography}
We seek to use the reputation management system EigenTrust~\cite{RF:eigen_trust_reputation_management} and threshold cryptography techniques along with Zero Trust Principles to build a robust trust algorithm and implement a combination of Eigentrust and ZTN on a sandbox, possibly starting from a totally centralized network controlled by a system administrator to a totally distributed network.  This step will be used after we have already recorded EDR data, applied tactical provenance analysis and applied further reduction of log records using Huffman coding. We may start first from an entirely centralized implementation as it will be easier to study the functionality and response of the system and then move on to a distributed version. EigenTrust is an algorithm that provides each peer in the network a unique global trust value based on the peer's history of uploads and thus aims to reduce the number of inauthentic file transactions in a Peer to Peer (P2P) network~\cite{RF:eigen_trust_reputation_management}. Recall that the EigenTrust algorithm is based on transitive trust, which means that if a peer $p$ trusts any peer $q$, then it would also trust the peers trusted by $q$. Each peer $p$ calculates the local trust value $S_{pq}$ for all peers that have provided it with authentic satisfactory or unsatisfactory transactions it had~\cite{RF:eigen_trust_reputation_management}. 
The following score is calculated for each peer combination~\cite{RF:eigen_trust_reputation_management}:
\[ S_{pq} = \mbox{sat} (p, q)  - \mbox{unsat} (p, q) \] 
The local value is then normalized~\cite{RF:eigen_trust_reputation_management}, to prevent malicious peers from assigning arbitrarily high local trust values to colluding malicious peers and arbitrarily low local trust values to good peers. Detailed information on reputation management can be found in~\cite{RF:eigen_trust_reputation_management}.

Threshold Cryptography protects information by encrypting it and distributing it among a distributed system of fault-tolerant computers~\cite{RF:threshold_crypto_1999},~\cite{RF:threshold_crypto_1994}. We encrypt a message using a public key, and the corresponding private key is shared among the participating computers. Any cryptography activity needs a yes gesture from a pre-decided number of participants equivalent to threshold as discussed in Section~\ref{sec:introduction}. Our final algorithm will seek benefits from methods and principles stated in this paper in order to ensure reduced implicit trust zones and provide security following Zero Trust principles.

\section{Discussions, Conclusions and Future Work}
\label{sec:conclusion}
Data from EDR can be deployed at various levels of the design of ZTA networks; however, keeping track of the entire EDR dataset is a complex process. Zero Trust Architecture is the need of the hour, keeping in view threats like Advanced Persistent Threats. For making this process simplified, reduced and more useful, we can use methods such as Tactical Provenance Analysis and Graph Reduction. Creation of a trust score algorithm should be flexible with the design, but rigid to ZTA principles. A generic way of evaluating trust vectors and computing confidence levels can be done using a combination of trust algorithms, reputation management algorithms and threshold cryptography. In the future, we seek to build trust algorithms using data that we record using EDR tools that can be adapted to changes in environment variables. One can think of more methodologies for log reduction such as using information theory for efficient storage and faster processing and determining possibilities of real time access control deployment. Another direction for future research would be a complete decentralization in ZTA, while keeping it real time. A flexibly coupled hardware and set of software can be used to build robust Zero Trust Architecture based systems for enterprises.


\end{document}